\documentclass[submission, Proceedings]{SciPost}

\begin{document}

\begin{center}{\Large \textbf{Neutrinoless Double Beta Decay Overview}}\end{center}

\begin{center}
L.~Cardani\textsuperscript{1*}
\end{center}

\begin{center}
{\bf 1} INFN - Sezione di Roma, Roma I-00185 - Italy
\\
laura.cardani@roma1.infn.it
\end{center}

\begin{center}
\today
\end{center}

\definecolor{palegray}{gray}{0.95}
\begin{center}
\colorbox{palegray}{
  \begin{tabular}{rr}
  \begin{minipage}{0.05\textwidth}
    \includegraphics[width=8mm]{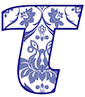}
  \end{minipage}
  &
  \begin{minipage}{0.82\textwidth}
    \begin{center}
    {\it Proceedings for the 15th International Workshop on Tau Lepton Physics,}\\
    {\it Amsterdam, The Netherlands, 24-28 September 2018} \\
    \href{https://scipost.org/SciPostPhysProc.1}{\small \sf scipost.org/SciPostPhysProc.Tau2018}\\
    \end{center}
  \end{minipage}
\end{tabular}
}
\end{center}


\section*{Abstract}
{\bf
Neutrinoless Double Beta Decay is a hypothesised nuclear process in which two neutrons simultaneously decay into protons with no neutrino emission. The prized observation of this decay would point to the existence of a process that violates a fundamental symmetry of the Standard Model of Particle Physics, and would allow to establish the nature of neutrinos. Today, the lower limits on the half-life of this process exceed 10$^{25}$-10$^{26}$ yr. I will review the current status of the searches for Double Beta Decay and the perspectives to enhance the experimental sensitivity in the next years.
}

\vspace{10pt}
\noindent\rule{\textwidth}{1pt}
\tableofcontents\thispagestyle{fancy}
\noindent\rule{\textwidth}{1pt}
\vspace{10pt}

\section{Introduction}
\label{sec:intro}
Described by Maria Goeppert-Mayer in 1935~\cite{DoubleBeta}, Double Beta Decay (DBD) is a transition between isobaric nuclei in which two neutrons simultaneously decay into protons. 
This transition is the rarest known nuclear decay as, being a second-order weak-interaction process, it is strongly suppressed and observable only for isotopes in which the single beta decay is forbidden. Despite its rarity, today it was observed for 11 nuclei with typical half-lives ranging from 10$^{18}$ to 10$^{24}$ yr~\cite{Barabash}.

The most interesting decay mode for DBD processes was discussed by Furry in 1937\cite{Furry} and is called \emph{neutrino-less} Double Beta Decay (0$\nu$DBD). In this hypothetical process, a nucleus would decay with the emission of two electrons but without the associated anti-neutrinos. 
The observation of this decay would be of the utmost importance for different Physics sectors. First of all, it would prove the existence of a process violating the conservation of the total lepton number. The reader may argue that this is only an accidental symmetry and thus its violation would not necessarily point to the Physics beyond the Standard Model. Nevertheless, 0$\nu$DBD would violate also the (B-L) quantity (baryon - lepton number) that, on the contrary, is a fundamental symmetry of the Standard Model. The detection of a process violating (B-L) would have important implications also in the theories that are trying to explain the asymmetry between matter and anti-matter in the Universe.
0$\nu$DBD has also fundamental implications in Neutrino Physics, as its existence would allow us to determine the nature of neutrinos. This process, indeed, can occur only if neutrinos, in contrast to all the other known fermions, are Majorana particles, i.e. they coincide with their own anti-particles. 
Many extensions of the Standard Model would lead to the existence of 0$\nu$DBD, but in this review I will focus on the light neutrino exchange. For a deeper discussion about other models we refer the reader to the comprehensive review reported in Ref~\cite{Dell'Oro:2016dbc}.

\section{A Challenge for Theorists and Experimentalists}
\label{sec:rarity}
The observable of 0$\nu$DBD is its half-life T$^{0\nu}_{1/2}$. Different experiments are measuring this parameter with increasing sensitivity and today current lower limits exceed 10$^{25}$-10$^{26}$ yr.
In the hypothesis that 0$\nu$DBD is mediated by light neutrino exchange, T$^{0\nu}_{1/2}$ can be related to a parameter containing the physics of neutrinos: the effective Majorana mass (m$_{\beta\beta}$):
\begin{equation}
T^{0\nu}_{1/2} = G^{0\nu}(Q,Z)\left | M^{0\nu} \right |^2 \left( \frac{ \left < m_{\beta\beta}\right >}{m_e}\right )^2.
\end{equation}
In this equation, G$^{0\nu}$ is the lepton phase-space integral, which depends on the Q-value of the decay and on the charge of the final state nucleus,
while M$^{0\nu}$ is the nuclear matrix element (NME), describing all the nuclear structure effects.
The effective Majorana mass  is defined as:
\begin{equation}
\label{MajoranaMass}
\left < m_{\beta\beta} \right > = \left | \sum_j m_j U_{ej}^2 \right | = \left | u_{e1}^2 e^{i\alpha_1} m_1 + u_{e2}^2 e^{i\alpha_2} m_2 + u_{e3}^2  m_3 \right |
\end{equation}
where $m_1$, $m_2$ and $m_3$ are the neutrino mass eigenvalues, u$_{e1}$, u$_{e2}$ and u$_{e3}$ are parameters of the Pontecorvo–Maki–Nakagawa–Sakata matrix describing neutrino oscillations, and $\alpha_1$, $\alpha_2$ two Majorana phases (meaningful only if neutrinos are Majorana particles).
Measurements of neutrino oscillations are not sensitive to the absolute values of $m_1$, $m_2$ and $m_3$, nor to the values of the Majorana phases. 
Nevertheless, they provide precise values concerning the square mass difference between the neutrino eigenstates, as well as the mixing angles~\cite{nufit,nufit2}. 
We can thus derive a range of possible values for m$_{\beta\beta}$ as a function of one of the neutrino masses (it is common to choose the lightest neutrino mass) or as function of the sum of the neutrino masses. Accounting for the current limits on m$_{\beta\beta}$ and the neutrino absolute masses, we expect m$_{\beta\beta}$ of 15-50 meV if neutrino masses were ordered according to the inverted hierarchy, and m$_{\beta\beta}<$5 meV if they followed the normal hierarchy~\cite{Dell'Oro:2016dbc}. The spread in the possible values is due to uncertainties in the parameters extracted from the oscillation measurements, and to the fact that the Majorana phases are unknown.

To convert these ranges into an expectation value for T$^{0\nu}_{1/2}$, we need G$^{0\nu}$ and the NME for each 0$\nu$DBD emitter.
G$^{0\nu}$ can be calculated very precisely, provided a sufficiently accurate description of the nuclear Coulomb effect on the emitted electron. For most of the emitters of interest, it is of the order of 10$^{-15}$-10$^{-16}$ yr$^{-1}$~\cite{G0nu}.

On the contrary, the NME range from 1 to 7 for most of the emitters, but deriving the precise value for each nucleus is still an open issue for theorists.
At present, a handful of groups are following different approaches: shell model \cite{ShellM1,ShellM2,ShellM3}, interacting boson model \cite{IBM}, different implementations of the quasiparticle random-phase approximation \cite{QRPA-1,QRPA-2,QRPA-3}, relativistic \cite{EFT-1,EFT-2} and non-relativistic  \cite{EFT-3} energy density functional theory. Different approaches result in a spread of about a factor 2-3 in the predicted value for the NME of a given isotope, suggesting that most of the models (if not all of them) are missing important elements.
Furthermore, some of the predictions of these models were compared to measurements, using the predictions for the single $\beta$ decay half-life (or for the double beta decay with two neutrino emission). To address the fact that the predicted half-lives are systematically faster than the measured ones, theorists made a phenomenological modification of the strength of the spin-isospin Gamow-Teller operator, by replacing the \emph{bare} axial coupling constant g$_A\sim$1.27 with an \emph{effective} value (g$_A^{eff}\sim$0.82\ g$_A$ for A$<$16, g$_A^{eff}\sim$0.77\ g$_A$ for 16$<$A$<$40, g$_A^{eff}\sim$0.74\ g$_A$ for 40$<$A$<$50 and g$_A^{eff}\sim$0.68\ g$_A$ for 60$<$A$<$80).
The uncertainty on the determination of g$_A$ clearly afflicts the determination of the NME but, since the source of this quenching is still unknown, it is difficult to predict the final effects on 0$\nu$DBD searches. 

In their interesting review~\cite{NME}, J.~Engel and J.~Men\`endez claim that recent improvements in the use of chiral effective field theory and of non-perturbative methods will produce reliable NME over the next five or so years.
However, we can use the typical values for G$^{0\nu}$ and the NME neglecting these uncertainties, in order to infer the expected order of magnitude of T$^{0\nu}_{1/2}$.
A simple calculation allows to estimate T$^{0\nu}_{1/2}\sim$10$^{26}$-10$^{28}$ yr if neutrinos masses follow the inverted hierarchy, and T$^{0\nu}_{1/2}>$10$^{28}$ yr if they are ordered according to the normal hierarchy.
In other words, even by monitoring a ton-scale detector we would expect a few signal events per yr (less than 1 event per yr in the normal hierarchy scenario).
The measurement of such a low signal rate is very challenging. 
Nevertheless, we expect the next-generation experiments to be able to cover part (if not the entire) region corresponding to the inverted hierarchy within the next decade.

\section{The Ideal Detector}
\label{sec:sensitivity}
Only a few isotopes can undergo 0$\nu$DBD.
From the theoretical point of view, for a given m$_{\beta\beta}$ we would expect a value T$^{0\nu}_{1/2}$ depending on the chosen isotope, thus the smartest choice would be the isotope with the shortest T$^{0\nu}_{1/2}$. Nevertheless, when combining the phase space factor and the NME, no isotope turns out to be particularly preferred over the others. 

\begin{figure}[!htbp]
\centering
\includegraphics[width=0.65\textwidth]{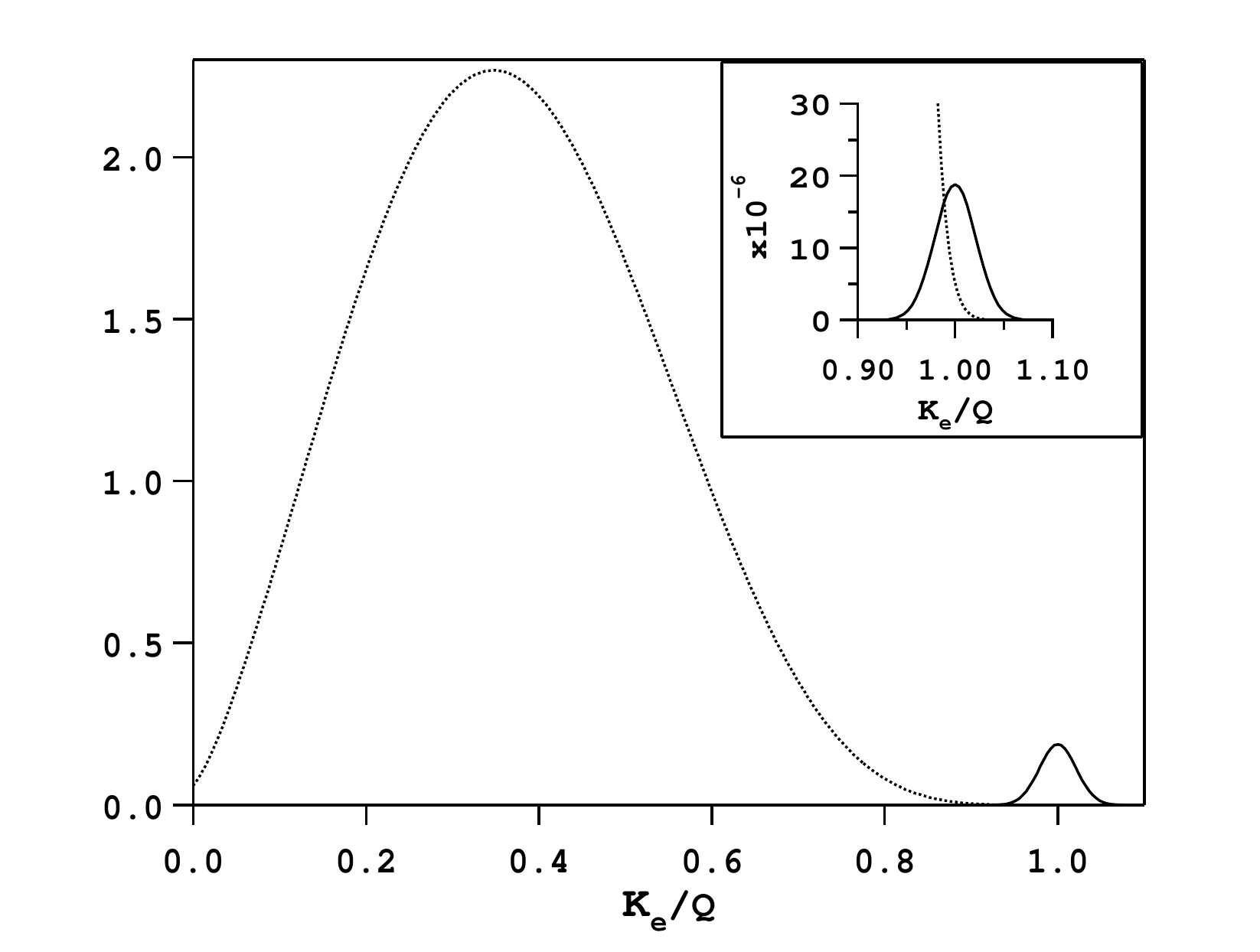}
\caption{Sum of the kinetic energies of the two electrons emitted in DBD. The dotted line from 0 to the Q-value is the distribution obtained in the 2-neutrino decay mode: since neutrinos carry away a variable fraction of the transition energy, the electrons show a continuous spectrum. The peak at the Q-value (not in scale) corresponds to a hypothetical 0$\nu$DBD signal. Inset: zoom around the Q-value.}
\label{fig:bb_spectra}
\end{figure}

The signature produced by a hypothetical 0$\nu$DBD is a monochromatic peak at the Q-value of the transition that, for most of the emitters, ranges from 2 to 3 MeV (Fig.~\ref{fig:bb_spectra}). 

The expected signature is thus very clear but, due to the rarity of the process, we have to pay attention to some features of the detector: 
\begin{itemize}
\item to fully explore the m$_{\beta\beta}$ region corresponding to the inverted hierarchy, we need a sensitivity of 10$^{26}$-10$^{28}$ yr. Thus, if we want to have a chance to observe at least a few decays, we need 10$^{26}$-10$^{28}$ atoms of the isotope of interest or, practically speaking, source masses of hundreds of kgs to tons.
\item Even if we can deploy such a large mass, we would expect only a few signal events. To have a discrete discovery potential, the background in the region of interest must be as close as possible to zero.
\item The 2$\nu$DBD is allowed by the Standard Model and, for any of the isotope that we choose, it will occur. We can select emitters with a long half-life for this decay mode, but it will still be orders of magnitude faster with respect to the searched signal. Since the tail of the 2$\nu$DBD extends up to the Q-value of the transition, where we search for 0$\nu$DBD, a detector with excellent energy resolution would be preferable. A good energy resolution would also suppress other sources of background in the RoI, including dangerous peaking background that could mimic the searched signal. 
\item 2$\nu$DBD could produce background in the RoI also via pile-up. For this reason, fast detectors would be preferred.
\end{itemize}
As of today, none of the technologies satisfies all these requirements and different collaborations are exploiting different strategies, reviewed in Sec~\ref{sec:experiments}.

\section{A common problem: background suppression}
\label{sec:bkg}
All the detectors searching for 0$\nu$DBD are located in deep underground laboratories, where the cosmic rays flux is strongly suppressed by the mountain overburden. 
Furthermore, they are usually equipped with passive or active shields to decrease also the background produced by residual muons and by the environmental radioactivity (rocks, air, ...).

When all the \emph{far} sources of background are suppressed, the detector itself can become the main problem. All the materials for the detector construction must be carefully selected, cleaned and stored to avoid possible re-contaminations.
These precautions unfortunately are not always sufficient to guarantee a negligible background. The residual contributions depend on the chosen technology and on its capability to reject interactions that differ from the electrons emitted by possible 0$\nu$DBD, such as $\alpha$ particles, $\gamma$ rays and electrons produced by radioactive contaminants.

As of today, most of the experiments can identify or reject $\alpha$ particles by exploiting the different time-development of pulses, or the event topology, or the light output. 

Electrons and $\gamma$ rays interact similarly in the detector and are thus more difficult to distinguish from electrons produced by 0$\nu$DBD. 
Some of the collaborations are working on the topological reconstruction to disentangle two electrons from a multi-Compton interactions or from events ascribed to single electrons.
More generally, it would be preferable to choose a 0$\nu$DBD emitter with high Q-value, as the $\beta/\gamma$ environmental background decreases at higher energies.
Under this point of view, there are two energies of particular interest: the 2.6 MeV line emitted by the decay of $^{208}$Tl, that is effectively the end-point of natural radioactivity, and the much rarer $\sim$3 MeV produced by $^{214}$Bi. Above 3 MeV, also this background source drops.

Isotopes like $^{40}$Ca, $^{96}$Zr and $^{150}$Nd are considered the golden candidates, as their Q-value lies well above 3 MeV and so they are less affected by background with respect to all the others. Unfortunately, these isotopes have also a very low natural isotopic abundance and their enrichment presents several technical difficulties. 
Emitters like $^{82}$Se, $^{100}$Mo and $^{116}$Cd feature a Q-value between 2.8 and 3 MeV, so they are barely affected by the $\beta/\gamma$ natural radio-activity. Also in this case the isotopic abundance is low (7-10$\%$), but the enrichment is feasible (though expensive).

Finally, there are isotopes with smaller Q-value, such as $^{130}$Te, and $^{136}$Xe that, nevertheless, feature a high natural abundance or can be easily enriched at low cost.

In summary, the choice of the isotope can give a substantial help in decreasing the background for those technologies lacking of tools for an active rejection of $\beta/\gamma$ events. Nevertheless, this choice comes at a cost of a low isotopic abundance. For this reason, different collaborations chose different approaches.

In the next Sections, I will briefly review the technologies that today reach the highest sensitivity for 0$\nu$DBD.   

\section{Experimental Status}
\label{sec:experiments}
Most of the 0$\nu$DBD experiments exploit a homogeneous approach, i.e., the detector coincides with the source, enhancing the efficiency for the collection of the electrons emitted in the decay.
A different approach consists of separating the source and the detector. The loss in efficiency is compensated by the better topological reconstruction of the single electrons, which could be very appealing in case of discovery.

The most advanced non-homogeneous detector was NEMO-3, operated between 2003-2011. In this experiment, different 0$\nu$DBD isotopes were deposited on thin foils for a total mass of 6.9 kg ($^{100}$Mo), 0.93 kg ($^{82}$Se), 0.45 kg ($^{130}$Te), 0.40 kg ($^{116}$Cd), 36.5 g ($^{150}$Nd), 9.43 g ($^{96}$Zr) and 6.99 g ($^{48}$Ca). The source foils were arranged in a cylindrical detector with an internal hole. The internal and external walls of the cylinder were equipped with calorimeter blocks, and the area between the foils and the calorimeter walls was filled with wire tracker cells.
The detector was equipped with several passive shields (iron, wood, borated water) and with an anti-radon tent.

The electrons emitted by the source foil followed curved trajectories (because of a magnetic field) and hit the calorimeters on the internal/external walls. The Geiger signal extracted from the drift cells allowed to reconstruct the tracks, while the calorimeters were used to reconstruct the total energy. 

NEMO-3 searched for many processes (see Refs.\cite{NEMO-3-Se,NEMO-3-Cd,NEMO-3-Nd,NEMO-3-Ca,NEMO-3-Mo,NEMO-3-quadruple} and references therein for the most recent results) and the interesting physics results motivated the endeavour of designing Super-NEMO~\cite{SuperNEMO}.
With the first demonstrator module in construction at the LSM underground lab in the Fr\'ejus tunnel (France), SuperNEMO aims at operating 7 kg of $^{82}$Se for 2.5 yr with a projected 90$\%$ C.L. sensitivity of T$_{1/2}^{0\nu}>$5.85$\times$10$^{24}$ yr. This value is not striking, considering that current experiments are probing half-lives in the range of 10$^{25}$-10$^{26}$ yr. Nevertheless, in case of discovery SuperNEMO would provide the unique possibility of reconstructing the tracks of the single electrons and thus distinguishing among different decay modes.
 
 In contrast to SuperNEMO, the experiments reviewed in the following sections are based on the ``homogeneous'' approach.
 
\subsection{Experiments searching for the 0$\nu$DBD of $^{136}$Xe}
Because of the low procurement cost and ease in enrichment, $^{136}$Xe was chosen by several collaborations.
KamLAND-Zen is leading the field in the search of the $^{136}$Xe 0$\nu$DBD, but also EXO-200 is reaching an interesting sensitivity, convincingly motivating the deployment of its successor, nEXO. In this section I briefly review also the NEXT project, that has not yet reached an interesting sensitivity on 0$\nu$DBD but is setting the basis for a technology that could have a very high discovery potential.

\subsubsection{KamLAND-Zen}
\label{sec:KamLAND}
As of today, the KamLAND-Zen experiment reached the most impressive limit on the half-life of 0$\nu$DBD. 
This experiment, located in the Kamioka mine (Japan) exploits the KamLAND (Kamioka Liquid scintillator Anti-Neutrino Detector) facility, in which $\sim$1000 tons of liquid scintillator were deployed in a 13 m diameter balloon. With a photo coverage of 30$\%$, Kamiokande features a FWHM energy resolution of about $\frac{15\%}{\sqrt{E\ [MeV]}}$ and a RMS vertex resolution of about $\frac{12\ cm}{\sqrt{E\ [MeV]}}$.

In the last years, the KamLAND detector was upgraded for 0$\nu$DBD searches, with the insertion of a mini-ballon containing liquid scintillator loaded with Xe 90.6$\%$ enriched in $^{136}$Xe (Q-value$\sim$2.458 MeV).
The detector started taking data in Sep 2011 using a mini-ballon with a radius of 1.54 m (320 kg of $^{136}$Xe). The fist project phase, completed in Jun 2012, allowed to collect an exposure of 89.5 kg$\cdot$yr and set a 90$\%$ C.L. lower limit of T$_{1/2}^{0\nu}>$1.9$\times$10$^{25}$ yr\cite{Kamland1}.
Unfortunately, this phase suffered from the presence of a peaking background in the region of interest. This contamination, observed along with $^{113}$Cs, was ascribed to $^{110m}$Ag coming from Fukushima reactor accident. For this reason, the KamLAND-Zen collaboration made a stop until Dec 2013 in order to purify the scintillator, and then restarted the data-taking until Oct 2015. Despite the presence of a contamination due to a pump failure, limiting the active volume to 43$\%$, this physics run allowed to reach a 90$\%$ C.L. limit of T$_{1/2}^{0\nu}>$9.2$\times$10$^{25}$ yr. 
The combined analysis of phaseI + phaseII data provides the most competitive limit on the 0$\nu$DBD half-life: T$_{1/2}^{0\nu}>$1.07$\times$10$^{26}$ yr, corresponding to m$_{\beta\beta}<$61-165 meV\cite{Kamland2}.

The beautiful result of KamLAND-Zen, that is the first experiment able to touch the region corresponding to the inverted hierarchy scenario, motivated the proposal of KamLAND-Zen 800, in which 750 kg of Xe gas will be inserted in a 1.92 m radius balloon with improved radio-purity with respect to the old balloon. The physics run of this experiment is foreseen to start in winter 2018 and aims at a sensitivity of T$_{1/2}^{0\nu}>$4.6$\times$10$^{26}$ yr.


\subsubsection{EXO-200 and nEXO}
\label{sec:EXO}
In contrast to KamLAND-Zen, the EXO-200 collaboration employs a liquid Time Projection Chamber (TPC) based on Xe. With a RMS energy resolution of 1.23$\%$/E and about 75 kg of Xe in the fiducial volume, EXO-200 reached a limit of T$_{1/2}^{0\nu}>$1.1$\times$10$^{25}$ yr in the first project phase\cite{EXO}. This collaboration was also able to study other interesting rare events, such as the Majoron-emitting modes, as well as Lorentz and CPT Violation in Double Beta Decay\cite{EXO1,EXO2}.

After an unfortunate incident at the Waste Isolation Pilot Plant, EXO-200 was forced to stop the physics runs. They started again in April 2016 with upgrades to the front-end electronics and Radon suppression system. This project phase (still on-going) allowed to set a preliminary limit of T$_{1/2}^{0\nu}>$1.8$\times$10$^{25}$ yr\cite{EXO3} and to improve the understanding of the background sources in view of a next-generation project: nEXO.

This ambitious experiment aims at operating 5000 kg of Xe enriched to 90$\%$ in $^{136}$Xe with improved energy resolution and lower background with respect to EXO-200 (thanks to the monolithic design). As explained in the pre-conceptual design review\cite{pCDR}, the final goal of nEXO will be the achievement of a sensitivity of the order of 10$^{28}$ yr.
It is worthy mentioning the recent progresses in the technology of Barium-tagging through Single Molecule Fluorescent Imaging\cite{BaTag} that would allow to tag the daughter of 0$\nu$DBD, thus dramatically reducing the background in view of a future (hypotetical) upgrade of nEXO.

\subsubsection{NEXT}
\label{sec:NEXT}
Another experiment searching for the 0$\nu$DBD of $^{136}$Xe is NEXT, which utilizes an electroluminescent high-pressure (10-15 bar) TPC based on Xe. 
The use of proportional electroluminescent amplification provides a large yield of photons per MeV, allowing to optimise the energy resolution: in contrast to liquid TPC and liquid scintillators, the NEXT prototypes proved a resolution better than 1$\%$ FWHM at the Q-value.
Furthermore, this technology provides the measurement of the topological signature of the event: the two electrons emitted in 0$\nu$DBD are expected to produce a track with constant energy deposition and two large clusters at the ends, which can be easily discriminated from different topological signatures produced, for example, by $\alpha$ particles or single electrons.

The R$\&$D of NEXT begun rather lately (2009) compared to other experiments, but the prototypes operated until now showed promising results.
The first mile-stone of the NEXT collaboration will be the operation of NEXT-100, foreseen for 2019 and aiming at T$_{1/2}^{0\nu}>$6.0$\times$10$^{25}$ yr in 3 yr\cite{NEXT1}.
Nevertheless, the LSC Scientific Committee recommended the installation of a demonstrator at the Underground Canfranc Laboratory, in order to validate the NEXT background model and to measure the 2$\nu$DBD to prove the potential of the topological signal. As a consequence, the collaboration decided to deploy the NEW (NEXT-WHITE) apparatus (scale 1:2 in size, or 1:8 in mass of NEXT-100).
Preliminary results obtained with this apparatus indicate a satisfactory radio-purity and an energy resolution of (0.4943$\pm$0.0017(stat)$\pm$0.0146(sys))$\%$  FWHM at the Q-value in the fiducial volume\cite{next-calib}.

\subsection{SNO+}
\label{sec:SNO}
SNO+ will exploit the facility that hosted the SNO experiment\cite{SNO} in VALE's Creighton mine near Sudbury (Ontario, Canada). 
The new detector is very similar to SNO: a 12 m diameter acrylic sphere filled with $\sim$800 tonnes of liquid scintillator floating in water, and equipped with about 10000 photomultipliers. In contrast to SNO, it was necessary to add ropes to hold the vessel down, since now it is filled with a light material (linear alkyl benzene scintillator). This facility will be multi-purpose, allowing the study of double beta decay, low energy solar neutrinos, reactor neutrinos, geo-neutrinos, supernova neutrinos and nucleon decay.

Concerning double beta decay, SNO+ decided to bet on the isotope $^{130}$Te, which stands out because of its natural isotopic abundance (about 33.8$\%$).
In Autumn 2019 the linear alkyl benzene scintillator will be loaded with 0.5$\%$ $^{nat}$Te, aiming at a sensitivity exceeding 10$^{26}$ yr in 1 yr, despite the rather poor energy resolution. A subsequent increase of the Te concentration (as well as the improvement of the light yield and the transparency) will allow to surpass 10$^{27}$ yr.
 
\subsection{LEGEND}
\label{sec:LEGEND}
The recently formed LEGEND collaboration is envisioning a next generation experiment based on High Purity Ge (HPGe) Diodes to search for the 0$\nu$DBD of $^{76}$Ge with sensitivity of the order of 10$^{28}$ yr\cite{LEGEND}.
The proposal of this experiment is based on the successful operation of two current-generation projects: GERDA (in the underground Laboratori Nazionali del Gran Sasso, Italy) and the Majorana Demonstrator (in the Sanford Underground Research Facility, South Dakota, USA).

In the GERmanium Detector Array (GERDA\cite{GERDA}) experiment, bare germanium detectors (with isotopic abundance in $^{76}$Ge ranging from 7.8 to 87$\%$) are operated in liquid argon, providing an active shield. Despite the low mass of tens of kg, GERDA was able to set one of the most competitive limits on the half-life of the $^{76}$Ge 0$\nu$DBD: 4.0$\times$10$^{25}$ yr in the first phase\cite{GERDA1}, increased to 8.0$\times$10$^{25}$ yr in the second one\cite{GERDA2}. 
The reasons at the basis of this success are the outstanding energy resolution (ranging from 2.7 to 4.3 keV FWHM at the Q-value) and the very low background level in the region of interest: 1.0$^{+0.6}_{-0.4}\times$10$^{-3}$ counts/keV/kg/y. This detector performance for other interesting searches, including possible signals produced by Physics beyond the Standard Model\cite{GERDA-Majorone,GERDA-2nu}.

The Majorana Demonstrator presents some differences with respect to GERDA, first of all the absence of the liquid argon active shield, that is replaced with high purity copper passive shields. With a total mass of 44.1 kg of Ge detectors (29.7 kg enriched in $^{76}$Ge), this experiment reached an unprecedented energy resolution of 2.52$\pm$0.08 keV FWHM at the Q-value and a slightly higher background in the RoI compared to GERDA: 6.7$^{+1.6}_{-1.4}\times$10$^{-3}$ counts/keV/kg/y considering the total statistics (despite a lower background was observed in the best experimental configuration). No 0$\nu$DBD candidates were observed in the first data release, leading to a limit of  1.9$\times$10$^{25}$ yr at 90$\%$ C.L.\cite{Majorana}. The collaboration exploited the detector also to search for other rare events, such as bosonic dark matter, solar axions and electron decay, as well as to probe the violation of the Pauli Exclusions Principle\cite{Majorana-rare}.

In the last year, members of the GERDA and the Majorana Demonstrator collaborations proposed a joint experiment, to be performed at the LNGS in the existing GERDA facility: LEGEND-200. To increase the sensitivity with respect to its predecessors, LEGEND-200 will exploit a larger mass (about 200 kg, in part re-using the existing GERDA and Majorana Demonstrator detectors) and a background reduced by a factor 3. Starting in 2021, LEGEND-200 aims at reaching a sensitivity larger than  10$^{27}$ yr with an exposure of 1 ton$\cdot$yr. This will be the basis for the tonne-scale experiment LEGEND.

\subsection{Cryogenic Calorimeters}
\label{sec:CUPID}
The technology of cryogenic calorimeters, originally proposed by Fiorini and Niinikoski\cite{Fiorini:1983yj}, consists of measuring the temperature variations induced by energy deposits in crystals operated at low temperatures (10 mK).
This technique offers many advantages: an exquisite energy resolution (0.1$\%$), high efficiency on the containment of the 0$\nu$DBD electrons and versatility, as most of the emitters of interest can be used as starting material to grow a crystal. This feature could be crucial in case of discovery, as it would allow to validate the result using another isotope.

CUORE (Cryogenic Underground Observatory for Rare Events\cite{Artusa:2014lgv}) is the most beautiful implementation of the cryogenic calorimeters technique. As in the case of SNO+, the CUORE collaboration chose $^{130}$Te. After the successful operation of the CUORE-0 demonstrator\cite{Alfonso:2015wka}, 206 kg of $^{130}$Te were embedded in 988 TeO$_2$ calorimeters (5$\times$5$\times$5 cm$^3$ each), then arranged in a mechanical structure made of high purity copper. The operation of CUORE was a real technological challenge, as it required the construction of a refrigerator able to cool 15 tons of material below 4 K, and the core of the detector (742 kg of TeO$_2$ and the copper structure) at 10 mK.
After a few weeks of data-taking the CUORE detector was stopped for an upgrade of the cryogenic facility. Despite the very short running time, CUORE already reached a competitive sensitivity of 1.5$\times$10$^{25}$ yr (90$\%$ C.L.)\cite{Alduino:2017ehq}, that will increase by about an order of magnitude in the next years.

Presently, the community of cryogenic calorimeters is envisioning the upgrade of CUORE: CUPID. The CUPID collaboration will face two challenges: increasing the active mass and decreasing the background in the RoI by two orders of magnitude\cite{Wang:2015taa,Wang:2015raa,Artusa:2014wnl}.
The idea is to reuse the CUORE cryogenic facility and increase the mass via isotopically enriched crystals.

Concerning the background reduction, the first milestone will be the suppression of the $\alpha$ interactions produced by contaminants located in the inert material of the detector. According to the CUORE experience, these interactions constitute the dominant source of background for experiments based on cryogenic calorimeters\cite{Alduino:2017qet}.
For this purpose, the CUPID collaboration will couple a light detector to each cryogenic calorimeter, in order to exploit the light yield for particle identification. In the last years several R$\&$D activities were performed, both using TeO$_2$ as in CUORE, and other crystals featuring a larger light emission (some of the most recent results can be found in Refs.\cite{Beeman:2013vda,Cardani:2013dia,Cardani:2013mja,Armengaud:2015hda,Artusa:2016maw,Bellini:2016lgg,Cardani:2018krv,Berge:2017nys} and references therein).
These efforts gave birth to two medium scale projects: Lumineu (Modane, France\cite{Armengaud:2017hit}) and CUPID-0 (LNGS, Italy\cite{Azzolini:2018tum,Azzolini:2018dyb,Azzolini:2018oph,Azzolini:2018yye}) to study the 0$\nu$DBD decay of $^{100}$Mo and $^{82}$Se respectively. Lumineu and CUPID-0 proved that the $\alpha$ background can be efficiently rejected and are now taking data to investigate sub-dominant background contributions in view of CUPID.

\subsection{Other promising experiments}
In this review I did not cover the projects that have not yet reached a competitive sensitivity. Nevertheless it is worthy mentioning the most promising ones.
\begin{itemize}
\item The \emph{CANDLES} experiment is searching for the 0$\nu$DBD of $^{48}$Ca using CaF$_2$ crystals. The main advantage of this project is the high Q-value of $^{48}$Ca (4272 keV), well above the contributions of the natural radioactivity, that comes at the cost of a very low natural isotopic abundance (0.187$\%$). The CaF$_2$ crystals scintillate in the UV-band, and can be easily grown with high transparency and purity (as CaF$_2$ is commonly used in commercial applications). They are immersed in a liquid scintillator tank acting as an active veto. Today, 305 kg of CaF$_2$ scintillators have been installed at the Kamioka laboratory to prove the background suppression. The preliminary analysis of these data provided a 90$\%$ C.L. lower limit on the half-life of $^{48}$Ca 0$\nu$DBD of T$_{1/2}^{0\nu}>$6.2$\times$10$^{22}$ yr\cite{CANDLES}.
\item \emph{COBRA} uses semiconductor detectors made of CdZnTe to search for the 0$\nu$DBD of  $^{114}$Cd, $^{128}$Te, $^{70}$Zn, $^{130}$Te, and $^{116}$Cd. Since the crystal growth starting from this material is not optimised as the crystal growth of more standard semiconductors, such as silicon or germanium, smaller crystals ($\sim$10 g) must be used to preserve good detection features. The final goal is the construction of a 3D array of small crystals (the final stage foresees 64k detectors)\cite{COBRA}.
\item The \emph{FLARES} collaboration is working on the enhancement of the collection of the scintillation light emitted by ultra-pure crystals, in order to reach an energy resolution of 2$\%$ with a large mass detector. For this purpose, they are exploiting a matrix of high performance silicon drift detectors cooled to 120 K\cite{FLARES1}. The most recent results of this project, comprising the first test of a CdWO$_4$ crystal coupled to an array of silicon drift detectors can be found in Ref.\cite{FLARES2}.
\item The China Dark Matter Experiment (\emph{CDEX}) is a relatively new germanium-based experiment to be deployed at the China's Jinping Underground Laboratory. Even if the primary goal is the search for Dark Matter, it will include a search for  0$\nu$DBD in the tonne-scale detector\cite{CDEX}.
\item \emph{PandaX-III} will exploit the double-phase Xenon TPC, originally developed for Dark Matter searches, also to study 0$\nu$DBD at the China Jin-Ping underground Laboratory. In the first phase, a high pressure gas TPC operated at 10 bar, will contain 200 kg, 90$\%$ enriched in $^{136}$Xe. According to the projected background and resolution, this detector will be able to reach a sensitivity of about 10$^{26}$ yr in three years of data-taking. A first 20 kg prototype already provided interesting results\cite{PANDA}.
\item \emph{AMoRE} (YangYang, Korea\cite{AMoRE}) is a proposed experiment to study the 0$\nu$DBD of $^{100}$Mo with CaMoO$_4$ enriched cryogenic calorimeters. The AMoRE collaboration is  operating a pilot experiment (with a total crystal mass of 1.9 kg) consisting of a number of commissioning runs to identify and suppress the sources of noise and background. A larger experiment with about 5 kg of crystals is in preparation. The success of this run will be crucial for the next phase, that foresees the deployment of about 200 kg of $^{100}$Mo.
\end{itemize}

\section{Conclusion}
The search for a rare process such as 0$\nu$DBD is very challenging both from the theoretical and the experimental point of view. Nevertheless, much progress has been made during the last years. 
More accurate models for the description of the 0$\nu$DBD allowed allowed a narrowing of the possible values for the nuclear matrix elements, reaching an uncertainty of about a factor 2-3 for most of the emitters. It is likely that this range will be further restricted in the next five or so years. 
From the experimental point of view, many collaborations proved the potential of their technologies reaching a sensitivity of 10$^{25}$-10$^{26}$ yr. Some of them already have a clear path for further upgrades that will allow them to start exploring the region corresponding to the inverted hierarchy of neutrino masses.





\nolinenumbers

\end{document}